\documentclass[onecolumn,aps,pra,noshowpacs,superscriptaddress]{revtex4}
\usepackage{epsf,epsfig,amsmath,amssymb,graphicx,times}
\begin{document}

\title{Heralded Source of Bright Multi-mode Mesoscopic Sub-Poissonian Light}

\author{T.~Sh.~Iskhakov}
\affiliation{Department of Physics, Technical University of Denmark, Fysikvej Building 309, Kgs. Lyngby 2800, Denmark}
\affiliation{Max-Planck Institute for the Science of Light, Guenther-Scharowsky-Str. 1 / Bau 24, Erlangen  D-91058, Germany}
\email{tiskha@gmail.com}
\author{V.~C.~Usenko}
\affiliation{Department of Optics, Palack\'y University, 17. listopadu 12, Olomouc 771 46, Czech Republic}
\author{U.~L.~Andersen}
\affiliation{Department of Physics, Technical University of Denmark, Fysikvej Building 309, Kgs. Lyngby 2800, Denmark}
\affiliation{Max-Planck Institute for the Science of Light, Guenther-Scharowsky-Str. 1 / Bau 24, Erlangen  D-91058, Germany}
\author{R.~Filip}
\affiliation{Department of Optics, Palack\'y University, 17. listopadu 12, Olomouc 771 46, Czech Republic}
\author{M.~V.~Chekhova}
\affiliation{Max-Planck Institute for the Science of Light, Guenther-Scharowsky-Str. 1 / Bau 24, Erlangen  D-91058, Germany}
\affiliation{University of Erlangen-N\"urnberg, Staudtstrasse 7/B2, Erlangen 91058, Germany}
\affiliation{Physics Department, Moscow State University \\ Leninskiye Gory 1-2, Moscow 119991, Russia}
\author{G.~Leuchs}
\affiliation{Max-Planck Institute for the Science of Light, Guenther-Scharowsky-Str. 1 / Bau 24, Erlangen  D-91058, Germany}
\affiliation{University of Erlangen-N\"urnberg, Staudtstrasse 7/B2, Erlangen 91058, Germany}




\begin{abstract}
In a direct detection scheme we observed 7.8 dB of twin-beam squeezing for multi-mode two-color squeezed vacuum generated via parametric down conversion. Applying post-selection, we conditionally prepared a sub-Poissonian state of light containing $6.3\cdot10^5$ photons per pulse on the average with the Fano factor $0.63\pm0.01$. The scheme can be considered as the heralded preparation of pulses with the mean energy varying between tens and hundreds of fJ and the uncertainty considerably below the shot-noise level. Such pulses can be used in metrology (for instance, for radiometers calibration) as well as for probing multi-mode nonlinear optical effects.
\end{abstract}


\maketitle


Quantum optics provides the recipes to overcome the noise limit set by the corpuscular nature of light, also known as a shot noise level~(SNL). At the SNL, completely uncorrelated particles follow a Poissonian distribution in their numbers. Light with noise properties suppressed below the SNL (squeezed light) plays an essential role in quantum metrology~\cite{Bri06}, quantum imaging~\cite{Bri10,Mar16}, quantum communication~\cite{Nie00,Mad12}, and can be helpful in bio-sensing~\cite{Tay13}. Significant interest in the multi-mode squeezed light stems from the possibility to improve the spatial resolution of optical images~\cite{KolobovPRL00,Kolobov99}. While single-mode sub-Poissionian light was successfully generated in a wealth of experiments using displaced squeezed vacuum states by means of feedforward~\cite{Mer90} and postelection~\cite{Fabre,Zou06} techniques, constant-current-driven semiconductor diodes~\cite{Tap87,Mac88}, conditional preparation of bright multi-mode light with suppressed noise properties has never been demonstrated. The reasons are twofold: (i) the mode mismatch~\cite{Ivan,Isk14,Fin15} in the detection, which always reduces the measured nonclassical correlation between the multi-mode mesoscopic twin-beams used for the heralded sub-Poissonian light preparation, and (ii) low quantum efficiency and high dark noise of the heralding detectors. While the second problem regards to the quality of equipment, the solution to the first problem is to work with the low-photon number states, for which the excess thermal fluctuations of the unmatched modes are negligible. Therefore, to date, the prepared multi-mode sub-Poissonian light contained at maximum only 12 photons per pulse ~\cite{Bondani, Perina13}. Because of the low efficiency of the detectors, the achieved suppression of the photon-number noise was only $15\%$ below the SNL. It is a very attractive goal to generate multi-mode mesoscopic nonclassical light that can be used to reduce photon-number noise in the applications (for example, in radiometry or quantum imaging) or to probe nonclassical effects in matter by seeding them with bright nonclassical states.

In this work, we address both issues listed above. We generate highly multi-mode twin beams containing up to 1.4 photons per mode and $6.3\cdot 10^5$ photons per pulse on the average and detect them separately by two high-efficiency low-noise photodetectors. We observed, to the best of our knowledge, the strongest nonclassical correlations between multi-mode signal and idler squeezed vacuum beams. Using the method developed in Ref.~\cite{Fabre}, we conditionally prepare bright multi-mode sub-Poissonian light with the noise suppressed $(37\pm1)\%$ below the SNL. Since the pulse contains at least $10^5$ photons, it is the strongest source of multi-mode sub-Poissonian nonclassical light. Differently to the sub-Poissonian light prepared by displacing the squeezed state, our method does not require the phase stability of the coherent displacement.

\begin{figure}[htbp]
\centering
\includegraphics[width=60mm]{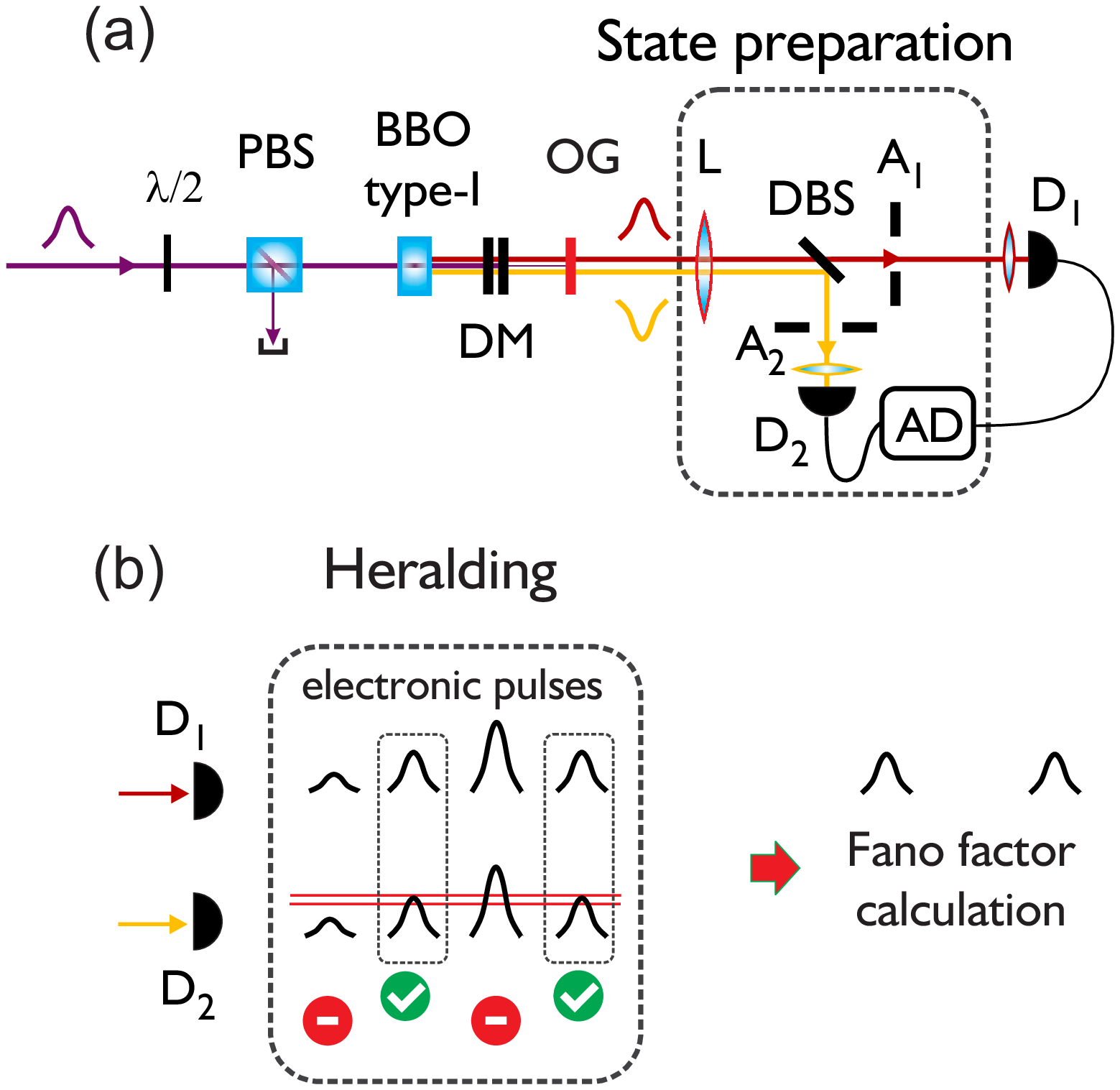}
\caption{(a) The experimental setup.  (b) Preparation of the Sub-Poissonian light through the heralding of the channel 2.}
\label{setup}
\end{figure}

The setup is shown in Fig.\ref{setup}~(a). The optical parametric amplifier OPA based on a single type-I 3-mm thick BBO crystal was pumped by a third harmonic of a Nd:YAG laser at the wavelength 355 nm with the pulse duration 18 ps and the repetition rate 1 kHz. The pump power was changed in the range from 36 mW to 126.3 mW by means of a half-wave plate ($\lambda/2$) placed in front of a polarization cube ($PBS$). The diameter of the pump beam (FWHM) at the position of the crystal was 1.5 mm. Signal and idler beams were generated at 635 nm and 805 nm, respectively. Right after the crystal the pump beam was cut off by two dichroic mirrors (DM) with high reflection at 355 nm and high transmission at 635 nm and 805 nm and a colored glass filter OG-630 (OG). The two-color parametric beams were separated by a dichroic beamsplitter (DBS). The apertures A1 and A2 of diameters 7 mm and 8.91 mm, respectively, were inserted into the signal and idler beams in the focal plane of a lens (L) with the focal distance 200 mm. The detected maximal angles satisfied the condition for the conjugate mode detection $\lambda_s/\lambda_i=\theta_s/\theta_i$ \cite{Ivan}. After the apertures, all the radiation was focused by two collecting lenses onto detectors ($D_1,D_2$). The propagation losses from the crystal to the detectors were measured separately and amounted to 7\%. According to the datasheet~\cite{Hamamatsu}, the quantum efficiency (QE) of the PIN diode S3072 at 635 nm is 82\% and QE of the PIN diode S3883 at 805 nm is 89\%. However, we expect the efficiency to be higher, because the protection glass windows in front of the diodes were removed. The signals from the diodes were amplified and shaped by low-noise charge sensitive preamplifiers (A250) and shaping amplifiers (A275) by Amptek. Pulses from the detectors were integrated by 12-Bit, 8-channel digitizer NI
PXI-5105 by National Instruments with the sample rate of 60 MS/s (AD). The data was stored in a computer for further processing.
The number of photons per mode was estimated from the nonlinear dependence of the PDC signal on the pump power~\cite{Iva06}. In the experiment the mean number of photons per mode $N_m$ of the parametric radiation was in the range from 0.3 to 1.4. For each mean value of the pump power, the data of $3\cdot 10^5$ signal and idler pulses were measured and stored in the computer.\\

To demonstrate the nonclassical photon-number correlations between the signal and idler beams using the data collected in the measurement we use the noise reduction factor (NRF), which is given by the formula
\begin{equation}
NRF\equiv \frac{Var(N_1-k\cdot N_2)}{Var(N_{1coh}-k\cdot N_{2coh})},
\label{NRF1}
\end{equation}
where $N_{1(2)}$ is the number of detected photons integrated per pulse in the detector $D_{1(2)}$ and $k=\langle N_1\rangle/\langle N_2\rangle$ is a numerical coefficient to compensate for the unbalancing of the detection gains (including detection efficiencies). We equalize the gains of the detectors to compensate for the excess noise caused by the thermal statistics of the detected beams~\cite{Brid10}. The value in the denominator is the shot-noise calibration for the given values $N_1,N_2$ and $k$ measured with the coherent light. If $NRF<1$ the photon-number correlation is nonclassical, i.e. it cannot be obtained by a mixture of classical coherent states. Taking into account the electronic noise of the detectors  the formula reads: 

\begin{equation}
NRF_{est}=\frac {Var(N_1-k\cdot N_2)-Var(V_{D1})-k^2Var(V_{D2})}{Var(N_{1coh}-k\cdot N_{2coh})-Var(V_{D1})-k^2Var(V_{D2})},
\label{NRF}
\end{equation}
where $V_{D1(2)}$ is the electronic and dark noise of detector $D_{1(2)}$. It should be noted that this equation can only be applied when the electronic noise of the detectors is smaller than the shot-noise level~\cite{Aga10}. This was the case in our experiment.

\begin{figure}[htbp]
\centering
\includegraphics[width=60mm]{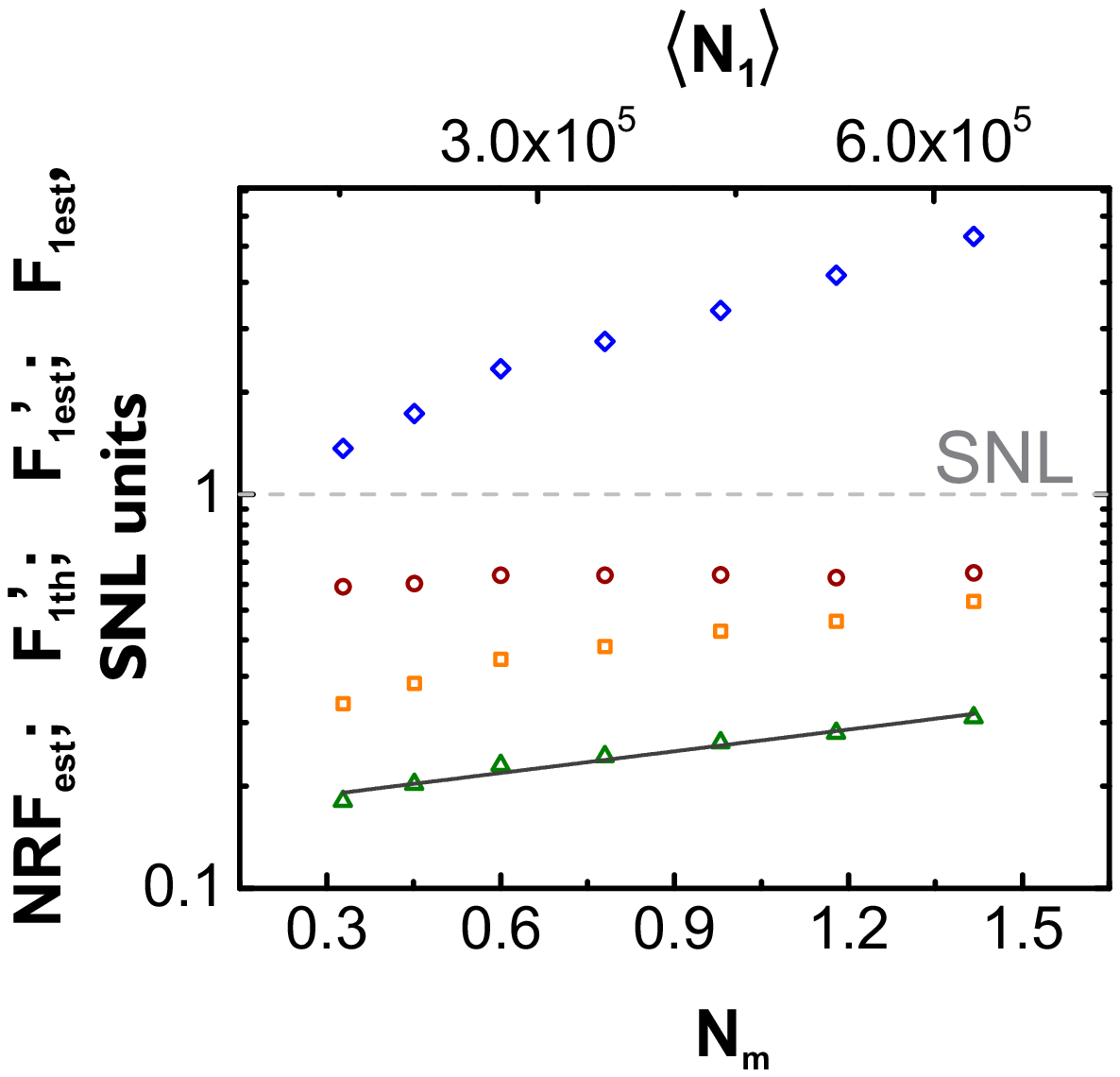}
\caption{Experimentally measured noise reduction factor $NRF_{est}$ (green triangles), theoretically predicted Fano factor $F_{1th}^{\prime}$ (orange squares) according to Eq.~(\ref{Fanoth}), the Fano factor $F_{1est}^{\prime}$ for the heralded state ($Q=20$) (vinous circles), and the Fano factor for the unconditional state $F_1est$ (blue diamonds) as functions of the number of photons per mode $N_{m}$ (bottom axis) and mean number of photons per pulse $\langle N_1\rangle$ (top axis). The straight line is a linear fit plotted according to Eq.~(\ref{NRFlin})}
\label{fig:2}
\end{figure}

As a first result, we plot NRF calculated according to Eq.~\ref{NRF} against the number of photons per mode in Figure.~\ref{fig:2}. As expected, we observed a linear dependence~\cite{Isk14}. The data were fitted by a linear function according to the formula
\begin{equation}
NRF_{est}=1-\alpha+ \beta \cdot N_{m},
\label{NRFlin}
\end{equation}
where the fitting parameters $\alpha=0.857\pm0.004$ and $\beta=0.0916\pm0.005$. Both parameters depend on the quantum efficiency of the first detector $\eta_1$, ratio of  quantum efficiencies of the detectors $k=\frac{\eta_1}{\eta_2}$, the number of matched modes $M$ and unmatched ones $K$ as $\alpha=\frac{2M}{M+K}\frac{\eta_1}{1+k}$ and $\beta=\frac{2K}{M+K}\frac{\eta_1}{1+k}$. Assuming that $M \gg K$ and the ratio of quantum efficiencies equals the ratio of the readings of the signal and idler detectors, which can be obtained from the measured data, we obtain $\eta_1=(86.2\pm0.5)\%$. As one can see from Figure \ref{fig:2}, we observed a reduction of the noise of the photon number difference down to $(16.6\pm0.3)\%$ of the SNL for the state containing $N_m=0.33$ and $\langle N_1\rangle=1.6\cdot10^5$. This amounts to 7.8 dB, the largest degree of twin-beam squeezing ever demonstrated for multi-mode squeezed vacuum states of light.

Conditional preparation of the sub-Poissonian light is realized using the same data set. The procedure of the data processing is based on post-selection. As shown in Fig.\ref{setup}~(b), for further analysis of detector $D_1$ data, we use only those pulses for which the output of detector $D_2$ takes  values within the range of $\frac{2 \cdot SD}{Q}$ around the chosen level. Here, $SD$ is the standard deviation of the signal and $Q$ is a constant defining the conditioning strength. The higher Q the stronger the condition.\\

To quantify the noise of the conditionally prepared light we calculate the Fano factor $F_{1(2)}\equiv \frac{Var(N_{1(2)})}{Var(N_{1(2)coh})}=\frac{Var(N_{1(2)})}{\langle N_{1(2)}\rangle}$. Following Ref.~\cite{Fabre04}, in our case, theoretical calculation of $F_{1th}^{\prime}$ for the conditionally prepared state in Channel 1 is given by
\begin{equation}
F_{1th}^{\prime}=F_1-\frac{(F_1+F_2-2NRF)^2}{4F_2},
\label{Fanoth}
\end{equation}
where $F_{1,2}$ are the Fano factors for the unconditional beams 1,2. Following this, we expect the noise in the prepared beam to be almost twice as large as the noise of the photon number difference.

Taking into account the electronic noise of the detector 1(2), the Fano factor reads
\begin{equation}
F_{1(2)est}=\frac{Var(N_{1(2)})-Var(V_{D1(2)})}{\langle N_{1(2)}\rangle-\langle V_{D1(2)}\rangle},
\label{FanoEq}
\end{equation}
Contrary to the NRF calculation (\ref{NRF}), in the Fano factor $F_{1est}^{\prime}$ for the heralded sub-Poissonian light in channel 1, only the noise of detector $D_1$ can be eliminated, but the electronic noise of the control detector (D2) is always present.

We next demonstrate the conditional preparation of the sub-Poissonian light. In Fig.~\ref{fig:2}, the Fano factor for conditionally prepared states ($Q=20$) in the target channel (vinous circles) as a function of $N_m$ is depicted. The points were calculated according to Eq.~(\ref{FanoEq}). The deviation of the experimental results from the theoretical predictions according to Eq.(\ref{Fanoth})  (orange squares) is due to the electronic noise contribution of the control detector (D2). For the brightest state containing $\langle N_1\rangle =6.3\cdot10^5$ photons per pulse the Fano factor is found to be $F_{1est}^{\prime}=0.63\pm0.01$, which is slightly larger than $F_{1th}^{\prime}=0.53\pm0.01$ predicted by Eq.~(\ref{Fanoth}) using $NRF_{est}=0.314\pm0.002$, $F_{1est}=4.53\pm0.01$, and $F_{2est}=4.33\pm0.01$.

\begin{figure} [htb]
\centerline{\includegraphics[width=10cm]{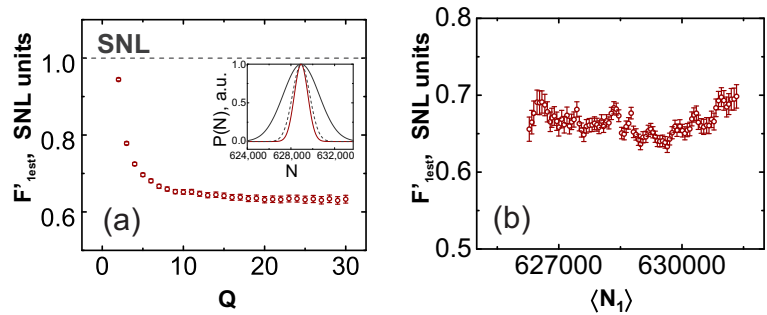}}
\caption{(a) The Fano factor plotted versus the strength $Q$ of the condition for the state containing $6.3\cdot10^5$ photons per pulse. Inset: probability distribution for the unconditional state (dark grey solid line), shot-noise limited state (light grey dashed line), and a conditionally prepared sub-Poisonian state (vinous solid line). (b) The Fano factor versus the mean number of photons $\langle N_1\rangle$ increased by shifting the middle of the conditioning interval with the bandwidth of $SD/10$ ($Q=20$) in channel 2.}
\label{FQN1}
\end{figure}

Apart from the electronic noise of the detectors, the signal and the idler channels are influenced by optical losses, finite quantum efficiency, and unmatched modes. Therefore, the selected interval in the control channel should have a finite width. For the brightest conditionally prepared state containing $6.3\cdot10^5$ photons per pulse, we plot $F_{1est}^{\prime}$ as a function of $Q$ in Fig.~\ref{FQN1}(a)). The conditioning was applied around the mean signal in $D_2$. As expected, the stronger the condition the lower the value of $F_{1est}^{\prime}$. According to the presented data, the Fano factor reaches the minimum value of $0.63\pm0.01$ at $Q=20$ and remains constant for $Q>20$. In general, the stronger the condition the lower the success rate of the state preparation. For example, applying the condition of widths $SD$, $SD/4$ and $SD/10$ we prepare sub-Poissonian light with the Fano factor of $F_{1est}^{\prime}(1)=0.944\pm0.004$, $F_{1est}^{\prime}(2)=0.66\pm0.01$ and  $F_{1est}^{\prime}(3)=0.63\pm0.01$ with the success rate of $40\%$, $10\%$ and $4\%$, respectively. The inset presents the probability distributions of the unconditional state (dark grey), the conditionally prepared sub-Poissonian state with $Q=20$ (vinous), and the shot noise limited state (light grey).

To demonstrate the stability of the method, for the target beam we plot $F_{1est}^{\prime}$ as a function of the mean number of photons per pulse (Fig.~\ref{FQN1}(b)). Each point of the plot was obtained from the same data set, measured at the pump power of 126.3 mW. The selection interval of bandwidth $SD/10$ ($Q=20$) was shifted around the mean signal in the control channel. One can see that the mean number of photons in the prepared state increases by $4\cdot10^3$ photons while the Fano factor remains almost unchanged within the error range.

In conclusion, we have measured a record value of 7.8 dB twin-beam squeezing for multi-mode two-color squeezed vacuum beams. Sub-Poissonian light containing up to $6.3\cdot10^5$ photons per pulse and up to $1.4$ photons per mode with the Fano factor $F_{1est}^{\prime}=0.63\pm0.01$ has been prepared in the target (signal) beam through the heralding of the control (idler) beam. The obtained pulses of light have the average number of photons from $1.5 \cdot 10^5$ to $6.3 \cdot 10^5$, corresponding to energies from $0.04$ to $0.15$ pJ, and the energy variance 37\% below the shot-noise level. This makes our technique interesting for radiometry, especially taking into account the possibility to move the idler beam wavelength to the infrared range. Besides, due to the relatively high photon number per mode, the obtained sub-Poissonian beam can be used now to probe the basic quantum interactions in nonlinear optics~\cite{Dem08} or quantum optomechanics~\cite{Van13} by mesoscopic nonclassical states of light.

We acknowledge the financial support of the EU FP7 under grant agreement No. 308803 (project BRISQ2). T.Sh.I. acknowledges the support of an H.C. \O{}rsted Postdoc programme, co-funded by Marie Curie Actions. R.F. acknowledges project GB14-36681G of Czech Science Foundation. V.C.U. acknowledges project 13-27533J of Czech Science Foundation.

\end{document}